# Epitaxial growth and anisotropy of La(O,F)FeAs thin films deposited by Pulsed Laser Deposition

E. Backen[1], S. Haindl[1], T. Niemeier[1], T. Freudenberg[1], J. Werner[2], G. Behr[2], L. Schultz[1] and B. Holzapfel[1]

[1]IFW Dresden, Institute for Metallic Materials, P.O. Box 270116, 01171 Dresden, Germany

[2]IFW Dresden, Institute for Solid State Research, P.O. Box 270116, 01171 Dresden, Germany

## Abstract

LaFeAsO$_{1-x}$F$_x$ thin films were deposited successfully on (001)-oriented LaAlO$_3$ and MgO substrates from stoichiometric LaFeAsO$_{1-x}$F$_x$ polycrystalline targets with fluorine concentrations up to $x = 0.25$ by PLD. Room temperature deposition and post annealing of the films yield nearly phase pure films with a pronounced c-axis texture and a strong biaxial in-plane orientation. Transport measurements show metallic resistance and onset of superconductivity at 11 K. $\mu_0 H_{c2}(T)$ was determined by resistive measurements and yield $\mu_0 H_{c2}$ values of 3 T at 3.6 K for $\mu_0 H \parallel c$ and 6 T at 6.4 K for $\mu_0 H \parallel ab$.

The recently discovered oxypnictide superconducting systems with Tc's up to 55 K have triggered intense research all over the world.[1-4] These layered materials are particularly interesting as they are assumed to be the first example of unconventional superconductivity with a very high transition temperature in a transition metal compound not based on a cuprate system. At present, however, the pairing symmetry as well as the pairing mechanism is under vivid debate. In the tetragonal parent compound, LaFeAsO, it was shown that with increasing fluorine doping a transition from an insulating compound to a metallic one with a $T_c$ of 26 K at x = 0.11 is observed.[1]

Currently, the investigation of polycrystalline bulk samples is the basis of most research activities in this field. To obtain a basic understanding of the mechanism of superconductivity in detail, however, single crystalline or at least highly textured samples are required. Intense work towards single crystal growth already resulted in first small single crystals of the size of several µm.[5,6] Epitaxial thin films of these compounds would in addition open the realization of a variety of very important experimental investigations and detailed anisotropy measurements. Additionally, the successful fabrication of thin films may also lead to applications of the new superconducting compounds. So far no successful superconducting thin film deposition reports were published, probably due to the high As gas pressures at elevated temperatures, a quite strong sensitivity of the superconducting properties with regard to the stoichiometry of the films and the demanding experimental realization due to process safety requirements. However, it was reported just recently that the non-superconducting parent LaFeAsO-phase has been obtained by Pulsed Laser Deposition and reactive solid phase epitaxy.[7]

Here we report on the successful synthesis and characterization of biaxially textured LaFeAsO$_{1-x}$F$_x$ thin films, which were prepared by Pulsed Laser Deposition from stoichiometric targets at room temperature with subsequent post annealing.

Polycrystalline targets of LaFeAsO$_{1-x}$F$_x$ (x = 0.1 and 0.25) were prepared from pure components by using a two-step solid state reaction method, similar to that described by Zhu et al.[8]

Due to the formation of a liquid phase at 1007°C, the 940°C annealing step was prolonged in comparison to Zhu et al.[8] to improve the homogeneity (for more details see Drechsler et al.[9]).

Film deposition was performed in a standard Pulsed Laser Deposition chamber, which was extended by a glove box based load lock system to enable the handling of the As-based compounds. A Lambda Physics laser (LPX305), operating with 248 nm KrF radiation at a laser frequency of 10 Hz was used to ablate polycrystalline $LaFeAsO_{1-x}F_x$ targets in high vacuum or Argon atmosphere up to 0.5 mbar. The target-substrate distance during film deposition was 4 cm and an ablation energy density of 4 J/cm$^2$ resulted in deposition rates of 0.7 nm/pulse under high vacuum deposition conditions of $10^{-6}$ to $10^{-5}$ mbar. (001)-oriented $LaAlO_3$ as well as MgO single crystals were used as substrates. The deposited total film thickness was usually about 600 nm. After film deposition at room temperature the films were sealed under vacuum into quartz tubes and annealed for various times at 1030°C. Basic structural analysis was done by x-ray diffraction in Bragg-Brentano geometry using a Philips X'Pert system operating with Co Kα radiation and by phi-scan measurements using a texture goniometer system operating with Cu Kα radiation. The chemical composition of the films was determined using a Philips XL series SEM with an additional Microspec WDX system and surface analysis was performed with a DI Atomic force microscope. A Quantum Design PPMS system equipped with a 9 T magnet was used for resistive measurements in a standard four-point geometry for field orientations $\mu_0 H \parallel c$ and $\mu_0 H \parallel ab$. A 90% resistance criterion was used for the $\mu_0 H_{c2}$ determination in both field directions.

$LaFeAsO_{1-x}F_x$ thin films were prepared under different deposition pressures from two different targets having a fluorine concentration of x = 0.1 and x = 0.25. WDX measurements of films prepared from the x = 0.1 fluorine target at 0.5 mbar Argon pressure showed reduced fluorine content and enhanced oxygen concentrations in the films compared to the target composition. Higher deposition pressures cause higher fluorine losses in the films due to enhanced scattering within the plasma plume. Due to this loss of fluorine during film deposition, targets with higher fluorine concentrations (x = 0.25) were prepared and used for laser ablation under high vacuum deposition conditions. During the heat treatment a significant As loss was detected for annealing temperatures above 1100°C.

Fig. 1 shows an AFM image of a 600 nm thick sample that was post annealed at 1030°C for 10 h. Growth islands with a typical size of 30 - 50 nm can be detected, which result in an overall surface RMS roughness of 79 nm.

Fig. 2 shows Co Kα x-ray diffraction patterns of a thin film deposited from a LaFeAsO$_{0.75}$F$_{0.25}$ target after subsequent post annealings for different times at 1030°C in a sealed quartz tube. The formation of the tetragonal c-axis oriented (00l) LaFeAsO$_{1-x}$F$_x$ phase with small (112) and (110) components can be observed. FeAs formation is pronounced after long annealing times and small amounts of secondary impurity phases like LaOF and La$_2$O$_3$ are also present in the films due to the high oxygen concentration as suggested by the WDX analysis.

A Cu Kα (102) phi-scan of the thin film deposited from a LaFeAsO$_{0.75}$F$_{0.25}$ target after 10 h annealing at 1030°C is shown in Fig. 3. A clear four-fold symmetry with FWHM values of ≈ 1° and a weak fibre texture can be observed in the film. A comparison with the (101) diffraction phi-scan of the underlying LaAlO$_3$ substrate (FWHM ≈ 0.5°) indicates an epitaxial growth relation of (001)[100] LaFeAsO$_{1-x}$F$_x$ || (001)[100] LaAlO$_3$ in spite of a lattice mismatch of 5.5%.

The results of the resistive measurements of two different films deposited from two different targets with fluorine concentrations of x = 0.1 on a MgO (001) substrate (a) and x = 0.25 on a LaAlO$_3$ (001) substrate (b) are shown in Fig. 4. The film deposited from the target containing only 10% of fluorine shows metallic behaviour down to T$_{min}$ ~ 75 K followed by a semiconducting increase in the resistance for lower temperatures and a small drop at T$_c$ ~ 3.6 K. The film was sealed in a quartz tube twice and annealed in two steps at 940°C for 1 h and at 1050°C for 1 h. The measurement was done after the second annealing step. Fig. 4(b) shows the R(T) curves for one sample deposited from the 25% fluorine target after three separated annealing steps. The sample shows nearly the same behaviour after the three annealing steps down to a temperature of T$_{min}$ ~ 60 K where a clear minimum can be observed followed by a small increase in resistance after the first two annealing steps. One can observe a superconducting transition like behaviour at $T_{c,1h} = 6.5$ K after 1h annealing and $T_{c,4h} = 11.1$ K after 4 h annealing. After the third annealing step (10 h) the minimum has vanished and the resistance decreases with a steeper slope below 60 K followed by a drop of the resistance at $T_{c,10h} = 9.6$ K. Although the resistance does not vanish completely, the R(T) curves show a strong hint for superconductivity since the drop in resistance shifts with increasing magnetic field towards smaller temperatures (see Fig. 5). This behaviour was already reported in previously prepared compounds and was explained by cracks in the sample and grain boundary contributions.[10,11] In the films presented here the detected quite high oxygen content results in minor non-superconducting secondary phases like $La_2O_3$ which might segregate at grain boundaries and therefore influence the resistance quite strongly, especially in the superconducting transition. The reduced $T_c$ of 11 K points to fluorine contents below $x = 0.1$ of the formed $LaFeAsO_{1-x}F_x$ phase as seen in polycrystalline bulk samples.[1,2] Nevertheless, the determination of the $\mu_0H_{c2}$ anisotropy of these epitaxially grown $LaFeAsO_{1-x}F_x$ films using electrical transport measurements in magnetic fields under different angles was possible.

Fig. 5 shows this $\mu_0H_{c2}$ evaluation for the film deposited from a $LaFeAsO_{0.75}F_{0.25}$ target after 10 h annealing at 1030°C (see Fig. 4 (b)). A 90% criterion was used to determine $\mu_0H_{c2}$ (see inset of Fig. 5) from resistive measurements and yields $\mu_0H_{c2}$ values of 3 T at 3.6 K for $\mu_0H \parallel c$ and 6 T at 6.4 K for $\mu_0H \parallel ab$. Extrapolation of the $\mu_0H_{c2}$ values to 6 K gives an anisotropy ratio of $\gamma = 7.8$ which is higher than values of $\gamma \sim 4.5$ determined by $NdFeAsO_{0.82}F_{0.12}$ single crystal measurements and $\gamma \sim 3.7$ in $LaFeAsO_{0.9}F_{0.1}$ polycrystalline bulk samples.[5,12]

Summarizing, Pulsed Laser Deposition was successfully used to deposit $LaFeAsO_{1-x}F_x$ thin films on (001)-oriented MgO and $LaAlO_3$ substrates. Post annealing at 1030°C in evacuated sealed quartz tubes results in nearly phase pure $LaFeAsO_{1-x}F_x$ films which grow epitaxially with a growth relationship of (001)[100] $LaFeAsO_{1-x}F_x \parallel$ (001)[100] $LaAlO_3$. Resistance measurements show superconducting transitions with a $T_{c,onset}$ of up to 11 K. Although the resistance does not vanish completely the magnetic field dependence of the resistance could be used for a $\mu_0H_{c2}$ anisotropy analysis. $\mu_0H_{c2}(T)$ values of 3 T at 3.6 K for $\mu_0H \parallel c$ and 6 T at 6.4 K for $\mu_0H \parallel ab$ were achieved. Reducing the oxygen excess in the films by using UHV deposition conditions and optimizing the annealing conditions should further improve the superconducting properties compared to polycrystalline samples which show Tc's up to 26 K.

ACKNOWLEGEMENT

The authors thank R. Hühne, M. Herrmann and K. Iida for the fruitful discussions and technical assistance.

FIG. 1. (Color online) Atomic force microscopy image of a thin film deposited from a LaFeAsO$_{0.75}$F$_{0.25}$ target after 10 h post annealing at 1030°C in a sealed quartz tube.

FIG. 2. Co Kα x-ray diffraction patterns of a thin film deposited from a LaFeAsO$_{0.75}$F$_{0.25}$ target after subsequent post annealings for different times at 1030°C in a sealed quartz tube. The formation of the (00l) oriented LaFeAsO$_{1-x}$F$_x$ (LOFFA) phase as well as minor secondary phases like LaOF, FeAs and La$_2$O$_3$ can be observed.

FIG. 3. (Color online) Cu Kα (102) phi-scan of the thin film deposited from a LaFeAsO$_{0.75}$F$_{0.25}$ target after 10 h annealing at 1030°C in a sealed quartz tube showing a four-fold symmetry.

FIG. 4. (Color online) R(T) curves for a film (a) deposited from a LaFeAsO$_{0.9}$F$_{0,10}$ target on MgO (001) after two annealing steps and (b) deposited from a LaFeAsO$_{0.75}$F$_{0.25}$ target on LaAlO$_3$ (001) after three annealing steps in a sealed quartz tube.

FIG. 5. (Color online) $\mu_0 H_{c2}$ evaluation for $\mu_0 H \parallel ab$ and $\mu_0 H \parallel c$ for a film deposited from a LaFeAsO$_{0.75}$F$_{0.25}$ target on LaAlO$_3$ (001) after the third annealing step in a sealed quartz tube. The inset shows R(T) curves for the $\mu_0 H \parallel ab$ orientation. A 90% criterion was used to determine $\mu_0 H_{c2}$.

Figure 1

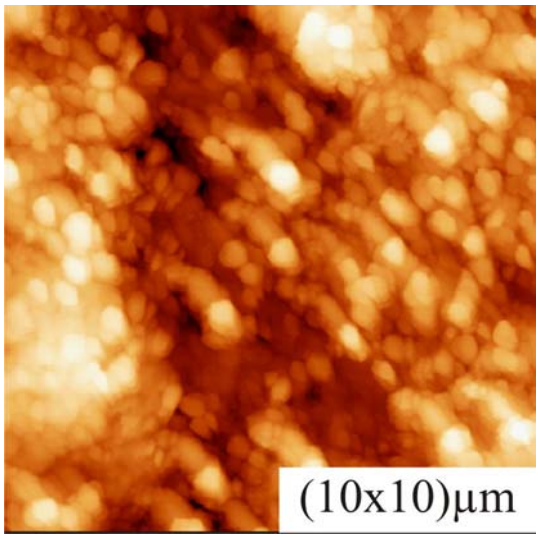

Figure 2

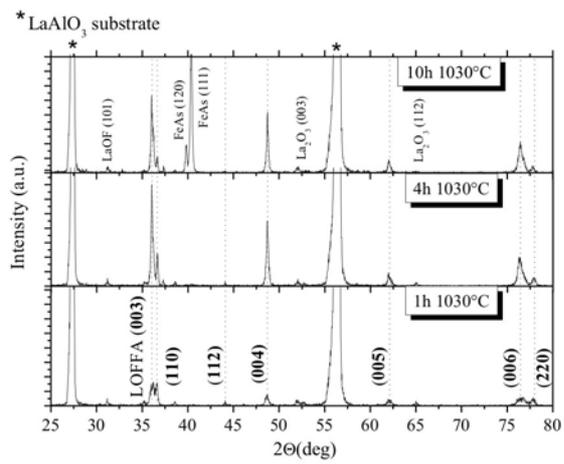

Figure 3

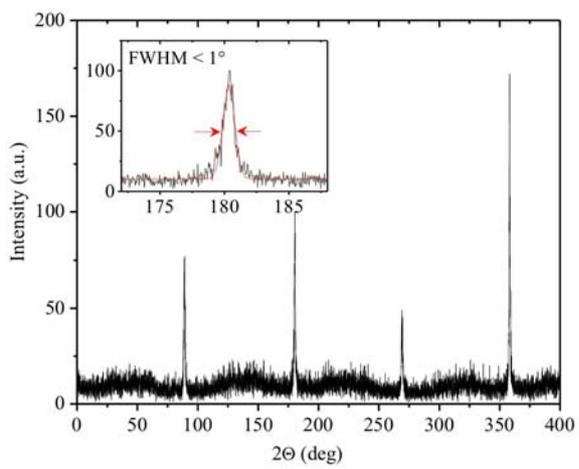

Figure 4

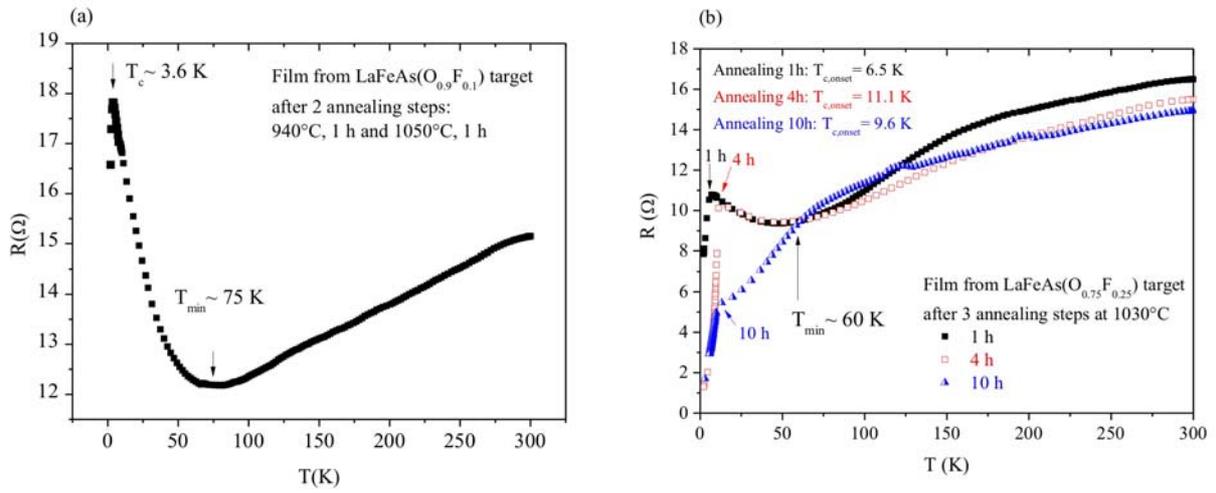

Figure 5

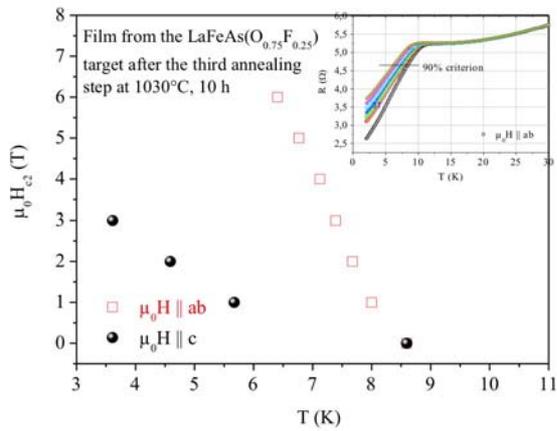